\documentclass[a4paper]{jpconf}
\usepackage{graphicx}
\def\Vec#1{\mbox{\boldmath $#1$}}
\def\kv{\mbox{\boldmath $ k$}}

\begin{document}
\title{Field angle dependence of the zero-energy density of states in unconventional superconductors: analysis of the borocarbide superconductor YNi$_2$B$_2$C}

\author{Yuki Nagai$^{1}$, Nobuhiko Hayashi$^{2,3}$, Yusuke Kato$^{1,4}$, Kunihiko Yamauchi$^{5}$ and Hisatomo Harima$^{6}$
}

\address{$^{1}$Department of Physics, University of Tokyo, Tokyo 113-0033, Japan}
\address{$^{2}$CCSE, Japan  Atomic Energy Agency, 6-9-3 Higashi-Ueno, Tokyo 110-0015, Japan}
\address{$^{3}$CREST(JST), 4-1-8 Honcho, Kawaguchi, Saitama 332-0012, Japan}
\address{$^{4}$Department of Basic Science, University of Tokyo, Tokyo 153-8902, Japan}
\address{$^{5}$CNR-INFM, CASTI Regional Lab, I-67010 Coppito (L'Aqulia), Italy}
\address{$^{6}$Department of Physics, Kobe University, Nada, Kobe 657-8501, Japan}


\begin{abstract}
We investigate the field-angle-dependent zero-energy density of states
for YNi$_2$B$_2$C with using realistic Fermi surfaces obtained by band calculations.
Both the 17th and 18th bands are taken into account.
For calculating the oscillating density of states,
we adopt the Kramer-Pesch approximation,
which is found to improve accuracy in the oscillation amplitude.
We show that superconducting gap structure determined by analyzing STM experiments is consistent with
thermal transport and heat capacity measurements.
\end{abstract}

The discovery of the nonmagnetic borocarbide superconductor
YNi$_2$B$_2$C \cite{Cava} has attracted considerable attention because
of the growing evidence for highly anisotropic superconducting
gap and high superconducting transition temperature 15.5K. 
In recent years,
Maki {\it et al}.\ \cite{Maki} theoretically suggested that the gap symmetry
of this material is $s$+$g$ wave and the gap function
has zero points (point nodes) in the momentum space. 
Motivated
by this prediction, field-angle dependence of the heat capacity \cite{Park}
and
the thermal conductivity \cite{Izawa}
have been measured on YNi$_{2}$B$_{2}$C.
The gap symmetry can be deduced from their oscillating behavior.
Those experimental results \cite{Park,Izawa} were considered to
be consistent with the $s$+$g$-wave gap. 
However, the present authors \cite{NagaiJ} recently found that 
the local density of states (LDOS) around a vortex
calculated for the $s$+$g$-wave gap 
on an isotropic Fermi surface (FS) 
is not consistent with
measurements by scanning tunneling microscopy and spectroscopy (STM/STS)~\cite{Nishimori}. 
Therefore, we calculated the LDOS around a vortex
with the use of a realistic FS of the 17th band obtained
by a band calculation \cite{NagaiY}.
We also investigated the density of states (DOS) under zero field and 
the field-angle dependence of the zero-energy DOS (ZEDOS).
Consequently, we proposed alternative gap structure for
YNi$_{2}$B$_{2}$C and succeeded in reproducing
those experimental observations consistently \cite{NagaiY}.
In this paper,
we investigate the field-angle-dependent ZEDOS taking the 18th band into account
in addition to the 17th band considered previously.
While the so-called Doppler-shift (DS) method was previously utilized in Ref.\ \cite{NagaiY},
we adopt here a more reliable method \cite{NagaiLett} for calculating the ZEDOS.

The DOS is the basis for analyzing physical quantities
such as the specific heat and the thermal conductivity \cite{Vorontsov}.
For example,
the specific heat $C/T$ is proportional to
the ZEDOS in the zero temperature limit $T \rightarrow 0$.
The DOS is obtained from the regular Green function $g$
within the quasiclassical theory of superconductivity,
which is represented
by a parametrization with $a$ and $b$ as
$g = - (1-a b)/(1+ a b)$.
They follow the Riccati equations ($\hbar=1$) \cite{Schopohl}: 
\begin{eqnarray}
\Vec{v}_{\rm F} \cdot \Vec{\nabla} a + 2 \tilde{\omega}_n a + a \Delta^{\ast} a - \Delta &=& 0,
\label{eq:ar}\\
\Vec{v}_{\rm F} \cdot \Vec{\nabla} b - 2 \tilde{\omega}_n b - b \Delta b + \Delta^{\ast} &=& 0.
\label{eq:br}
\end{eqnarray}
Here, $\Vec{v}_{\rm F}$ is the Fermi velocity, 
and $i \tilde{\omega}_n = i \omega_n + (e/c) \Vec{v}_{\rm F} \cdot \Vec{A}$ with
the Matsubara frequency $\omega_n$ and the vector potential $\Vec{A}$.

To analyze the field-angle-dependent experiments quantitatively, 
we have developed a method on the basis of the Kramer-Pesch approximation (KPA) \cite{NagaiLett}.
This method enables us to take account of the vortex-core contribution which is neglected
in the DS method.
By virtue of it, one can achieve quantitative accuracy.
We consider a single vortex situated at the origin of the coordinates.
To take account of the contributions of a vortex core,
in the KPA
we expand the Riccati equations (\ref{eq:ar}) and (\ref{eq:br})
up to first order in the impact parameter $y$ around a vortex
and the energy $\omega_n$~\cite{NagaiJ}
($i\omega_n \to E+i\eta$ and
$y$ is the coordinate along the direction perpendicular to $\Vec{v}_{\rm F}$). 
By this expansion,
one can obtain an analytic solution of the Riccati equations around a vortex. 
We then find the expression for the angular-resolved DOS \cite{NagaiLett}
\begin{equation}
N(E, \alpha_{\rm M}, \theta_{\rm M})
=
\frac{v_{\rm F0} \eta}
     {2 \pi^2 \xi_0}
\Biggl{\langle} 
\int 
\frac{ d S_{\rm F}  }
     { |\Vec{v}_{\rm F}|  }
\frac{
  \lambda
  \bigl[ \cosh
        (x/\xi_0)
  \bigr]^{\frac{-2 \lambda}{\pi h}} 
 }{(E-E_y)^2 + \eta^2} 
\Biggl{\rangle}_{\rm SP}.
\label{eq:dos}
\end{equation}
Here,
the azimuthal (polar) angle of the magnetic field $\Vec{H}$ is $\alpha_{\rm M}$ ($\theta_{\rm M}$)
in a spherical coordinate frame fixed to crystal axes,
and
$d S_{\rm F}$ is an area element on the FS
[e.g.,
$d S_{\rm F}=k_{\rm F}^2 \sin\theta d\phi d\theta$
for a spherical FS
in the spherical coordinates $(k,\phi,\theta)$,
and
$d S_{\rm F}=k_{{\rm F}ab} d\phi dk_c$
for a cylindrical FS
in the cylindrical coordinates $(k_{ab},\phi,k_c)$].
In the cylindrical coordinate frame $(r,\alpha,z)$ with
${\hat z} \parallel \Vec{ H}$ in the real space,
the pair potential is $\Delta \equiv \Delta_0 \Lambda(\kv_{\rm F}) \tanh(r/\xi_0) \exp(i\alpha)$
around a vortex,
$\Delta_0$
is the maximum pair amplitude in the bulk, $\lambda = |\Lambda|$,
and
$\langle \cdots \rangle_{\rm SP}
\equiv \int_0^{r_a} r dr \int_0^{2 \pi} \cdots d \alpha/ (\pi r_a^2)$
is the real-space average around a vortex,
where $r_a/\xi_0 = \sqrt{H_{c2}/H}$
[$H_{c2} \equiv \Phi_0 / (\pi \xi_0)$, $\Phi_0 = \pi r_a^2 H$].
$x=r\cos(\alpha -\theta_v)$,
$y=r\sin(\alpha -\theta_v)$,
and
$E_y = \Delta_0 \lambda^2 y/(\xi_0 h)$.
$\theta_v(\kv_{\rm F},\alpha_{\rm M}, \theta_{\rm M})$ 
is the angle of $\Vec{v}_{\rm F \perp}$ in the plane of $z=0$,
where $\alpha$ and $\theta_v$ are measured from a common axis \cite{NagaiJ,NagaiY}.
$\Vec{v}_{\rm F \perp}$
is the vector component of $\Vec{v}_{\rm F} (\kv_{\rm F})$
projected onto the plane normal to
${\hat {\Vec{H}}}=(\alpha_{\rm M}, \theta_{\rm M})$.
$|\Vec{v}_{\rm F \perp}(\kv_{\rm F},\alpha_{\rm M}, \theta_{\rm M})|
\equiv v_{\rm F0}(\alpha_{\rm M}, \theta_{\rm M}) h(\kv_{\rm F},\alpha_{\rm M}, \theta_{\rm M})$
and 
$v_{\rm F0}$ is the FS average of $|\Vec{v}_{\rm F \perp}|$ \cite{NagaiY}.
$\xi_0$ is defined as
$\xi_0 = v_{{\rm F}0}/(\pi \Delta_0)$.
We consider here a clean SC in the type-II limit.
The impurity effect can be incorporated through the smearing factor $\eta$.

We calculate the angular dependence of the ZEDOS
($E=0$ in Eq.\ (\ref{eq:dos}))
for YNi$_{2}$B$_{2}$C 
with using the band structure calculated by Yamauchi {\it et al}.\ \cite{Yamauchi}.
When integrating Eq.\ (\ref{eq:dos}),
the band structure is reflected in
$d S_{\rm F}$ and $\Vec{v}_{\rm F}$
($\Vec{v}_{\rm F \perp}$, $v_{\rm F0}$, and $h$ are obtained from $\Vec{v}_{\rm F}$).
In YNi$_{2}$B$_{2}$C there are three bands crossing the Fermi level, which are called
the 17th, 18th, and 19th bands.
The electrons of the 17th band predominantly contribute to the superconductivity,  
since the DOS at the Fermi level for the 17th, 18th, and 19th bands 
are 48.64, 7.88, and 0.38 states/Ry, respectively \cite{Yamauchi}.
We consider here the FSs of the 17th and 18th bands and neglect the 19th band
because the DOS is substantially small on the 19th FS.
We use an anisotropic $s$-wave gap structure obtained in our preceding paper
(Eq.\ (27) in Ref.\ \cite{NagaiY})
for the 17th FS.
This gap structure was determined by analyzing STM observations \cite{Nishimori}.
As for the 18th FS, we assume an isotropic gap with an amplitude
equal to the maximum gap on the 17th FS,
according to angular-resolved photo emission measurements \cite{Baba}.

\begin{figure}[h]
\begin{minipage}{10pc}
\includegraphics[width=14pc]{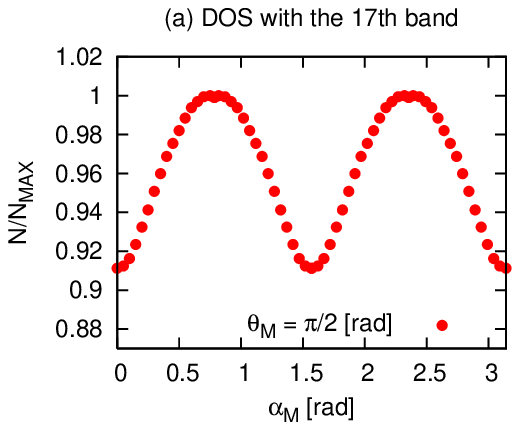}
\end{minipage}\hspace{2pc}%
\begin{minipage}{10pc}
\includegraphics[width=14pc]{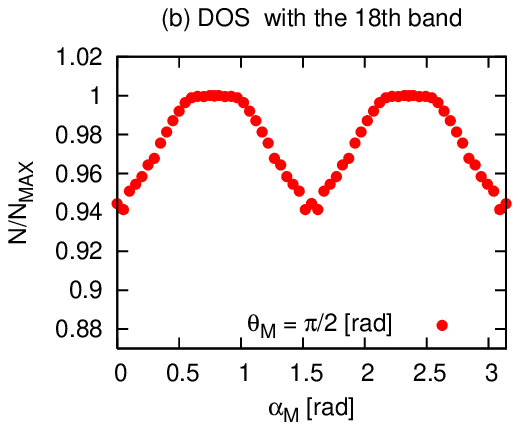}
\end{minipage} \hspace{2pc}%
\begin{minipage}{10pc}
\includegraphics[width=14pc]{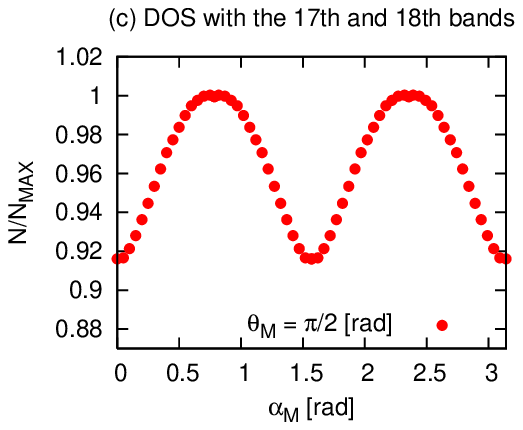}
\end{minipage} 
\caption{\label{fig:1}Angular dependence of the ZEDOS for YNi$_{2}$B$_{2}$C  (a):with use of the 17th band, (b): 
that of 18th band and (c) that of the 17th and 18th bands. 
The magnetic field tilts from the $c$ axis by polar angle $\theta_{M}=\pi/2$. 
$\eta = 0.05 \Delta_{0}$ and $r_{a} = 7 \xi_{0}$.}
\end{figure}

We rotate the applied magnetic field in the basal plane perpendicular to the $c$ axis
($\theta_{\rm M}=\pi/2$)
and investigate the azimuthal angle $\alpha_{\rm M}$ dependence of the ZEDOS. 
First, we show the partial ZEDOS on the 17th FS only. 
As seen in Fig.\ \ref{fig:1}(a), the oscillation amplitude is of the order 9~\%. 
Previously we have calculated the ZEDOS by the DS method under the 
same condition and obtained the amplitude of the order 30~\% \cite{NagaiY},
which was too large in comparison to experimental results \cite{Park,Izawa}.
The KPA adopted in this paper yields substantial improvement in the amplitude.

Second, we show the partial ZEDOS on the 18th FS only. 
As seen in Fig.\ \ref{fig:1}(b), the cusp-like minima appear. 
The origin of the cusp-like structure is different from that of the 17th-FS case, 
since the gap is isotropic on the 18th FS while it is anisotropic on the 17th FS. 
The cusp-like minima are due to the FS anisotropy on the 18th FS. 
According to the band calculation (Fig.\ 2(b) in Ref. \cite{Yamauchi}),
the shape of the 18th FS is like a square as viewed from the $k_{c}$ axis in the momentum space.
The sides of this square are parallel to the $k_{a}$ or $k_{b}$ axis.
Therefore, the region where the Fermi velocity is parallel to the $k_{a}$ or $k_{b}$ axis
has a high proportion of the area on the 18th FS.
Now, the quasiparticles with the Fermi velocity parallel to the magnetic field
do not contribute to the ZEDOS (e.g., see Ref.\ \cite{NagaiLett}).
As a result, when the magnetic field is rotated near the directions parallel to those axes,
the change in the ZEDOS is drastic, leading to cusp-like minima.
Udagawa {\it et al}.\ \cite{Udagawa} previously obtained
cusp-like minima in the field-angle-dependent ZEDOS by using a square-like FS model.
The origin of their cusp-like minima is the same as that of the present result for the 18th FS.
It should be noted that the DOS ratio $r$ of the 18th FS to the 17th FS
is $r = N_{\rm F 18}/N_{\rm F 17} = 7.88/48.64 \sim 0.16$ \cite{Yamauchi}.
The contribution of the 18th FS is not dominant,
and therefore the cusp-like minima observed in the experiments \cite{Park,Izawa}
probably cannot be attributed to the square shape of the FS only.

Finally, we show the total ZEDOS on both the 17th and 18th FSs.
As seen in Fig.\ \ref{fig:1}(c), the oscillation amplitude is of the order 8 \% and 
the overall behavior is almost similar to that for the 17th FS shown in Fig.\ \ref{fig:1}(a).
In the thermal conductivity measurements in YNi$_{2}$B$_{2}$C, 
the oscillation amplitude in the field 1 T and 0.5 T is of the order 2~\% \cite{Izawa}
and 4.5~\% \cite{Kamada}, respectively, at the same temperature 0.43 K. 
In the heat capacity measurements \cite{Park}, 
the oscillation amplitude in 1 T at 2 K is of the order 4.7~\%. 
The KPA is an approximation appropriate
in the limit of the low temperature and low magnetic field.
Therefore, in terms of the order of the oscillation amplitude,
our result could be considered to be consistent with that of those experiments.
As for the directions of the minima, the result is also consistent with the experiments.
Those mean that the assumed gap structure determined by analyzing STM experiments
(Eq.\ (27) in Ref.\ \cite{NagaiY})
is consistent with
the heat capacity \cite{Park} and the thermal transport \cite{Izawa} observations.
So far we have found here that inclusion of the 18th-FS contribution does not drastically alter
our previous conclusion \cite{NagaiY}, and
the use of the KPA improves the value of the amplitude
compared to our result \cite{NagaiY} by the DS method.
We should note, however, that the sharpness of the cusp is rather weak in Fig.\ \ref{fig:1}(c)
in comparison to the experimental observations \cite{Park,Izawa}.
This point may be resolved by considering carefully the $k_{c}$ dependence of the gap structure.
The gap structure used here was determined to explain
the azimuthal dependence of the observations in the basal plane \cite{NagaiY}.
A modification of the gap structure would be necessary to explain
the observed polar-angle dependence \cite{Izawa}.
This is left for future studies.
Here, it should be noted that
such a detailed discussion is almost impossible
till the KPA \cite{NagaiLett} is developed as a new method with high precision.

In conclusion, we calculated the field-angle-dependent ZEDOS on the FSs from 17th and 18th band
of YNi$_2$B$_2$C.
We adopted the gap structure  \cite{NagaiY} determined by comparison with
STM observations \cite{Nishimori}.
The oscillation amplitude was found to be of the order 8 \%.
This amplitude does not deviate so much from the heat capacity \cite{Park}
and the thermal transport \cite{Izawa,Kamada} measurements.
As for the directions of the minima, the result is also consistent with those experiments. 
This suggests that the assumed gap structure on the 17th FS is essentially appropriate
for the in-plane properties.
Further modification of the gap structure is expected in terms of the $k_c$ dependence.
The use of the KPA improved the value of the amplitude compared to
our previous result \cite{NagaiY} by the DS method. 
The KPA \cite{NagaiLett} appears to be an efficient method with high precision for analyzing
experimental data.

This research was supported by Grant-in-Aid for JSPS Fellows (204840),
and was also partially supported by the Ministry of Education, 
Science, Sports and Culture, Grant-in-Aid for 
Scientific Research on Priority Areas, 20029007.

\end{document}